\documentstyle[12pt,epsf]{article}
\newcommand {\beq}{\begin{equation}}
\newcommand {\eeq}{\end{equation}}
\newcommand {\beqa}{\begin{eqnarray}}
\newcommand {\eeqa}{\end{eqnarray}}
\newcommand {\beqan}{\begin{eqnarray*}}
\newcommand {\eeqan}{\end{eqnarray*}}
\newcommand {\n}{\nonumber \\}

\newcommand {\Eq}[1]{Eq.~(\ref{#1})}

\newcommand {\Romannumeral}[1]{\uppercase\expandafter{\romannumeral#1}}

\newcommand {\lsim} {\begin{array}{c} < \\[-3.2mm] \sim \end{array}}
\newcommand {\gsim} {\begin{array}{c} > \\[-3.2mm] \sim \end{array}}

\newcommand {\del}{\partial}
\newcommand {\eV}{\:{\rm eV}}

\newcommand {\GeV}{\:{\rm GeV}}
\newcommand {\TeV}{\:{\rm TeV}}

\newcommand {\PLB}[3]   {Phys. Lett.~{\bf B#1} (#2) #3}

\newcommand {\NPB}[3]   {Nucl. Phys.~{\bf B#1} (#2) #3}

\newcommand {\PRD}[3]   {Phys. Rev.~{\bf D#1} (#2) #3}
\newcommand {\PTP}[3]   {Prog. Theor. Phys. {\bf#1} (#2) #3}
\newcommand {\PP}[3]    {Phys. Rep.~{\bf#1} (#2) #3}
\newcommand {\PRL}[3]   {Phys. Rev.~Lett.~{\bf#1} (#2) #3}
 
\newcommand {\IJMPA}[3] {Int. J. Mod. Phys.{\bf A#1} (#2) #3}   
\newcommand {\MPLA}[3]   {Mod. Phys. Lett. {\bf A#1}(#2) #3} 

\begin{document}
\setlength{\oddsidemargin}{0cm}
\setlength{\baselineskip}{7mm}  

\begin{titlepage}
 \renewcommand{\thefootnote}{\fnsymbol{footnote}}
    \begin{normalsize}
     \begin{flushright}
                 KEK-TH-534\\
                 hep-ph/9708376\\
                 August 1997
~~\\
~~\\
~~\\
~~
     \end{flushright}
    \end{normalsize}
\vspace*{0cm}
    \begin{Large}
       \vspace{2cm}
       \begin{center}
         {\Large Baryogenesis at the Electroweak Scale and Above}
\footnote{To appear in the Proceeding of the KEK meeting on 
          ``$CP$ violation and its origin''.}     \\
       \end{center}
    \end{Large}

  \vspace{1cm}

\begin{center}
           Hajime A{\sc oki}\footnote
           {
e-mail address : haoki@theory.kek.jp,
{}~JSPS research fellow.}\\
      \vspace{1cm}
         {\it Theory Group, KEK,}\\
         {\it Tsukuba, Ibaraki 305, Japan} \\
\end{center}

\vfill

\begin{abstract}
\noindent 
We consider origins of the baryon asymmetry which we observe today.
We review the progress of electroweak-scale baryogenesis,
and show a new mechanism, string-scale baryogenesis.
\end{abstract}
\vfill
\end{titlepage}
\vfil\eject


\section{Introduction}\label{sec:intro}
\setcounter{equation}{0}
Baryogenesis \cite{Kolb Turner} is one of the important problems 
in particle physics  and cosmology:
Why do we live in a universe where there is dominance of  matter over 
anti-matter?
How do we explain the observed value of the ratio $n_B /s \sim 10^{-10}$,
where $n_{B}$ is the difference between the number density of baryons 
and that of anti-baryons, and $s$ is the entropy density?
Until now, many baryogenesis scenarios have been proposed,
where a nonzero $n_{B}$ is dynamically generated from  a baryon-symmetric 
initial state 
within the context of standard hot big bang cosmology.
 
In the theory of electroweak interactions (the Weinberg-Salam theory),
baryon number conservation is violated 
{\em via} sphaleron processes due to chiral anomaly.
Baryogenesis scenarios at the electroweak scale 
have been studied recently \cite{KRS,CKN,Rub Shap}.
They are attractive since they can be tested at the current and near-future 
terrestrial experiments.
Within the minimal standard model (MSM), 
it is difficult to generate the observed baryon asymmetry,
due to the too small $CP$ violation source and the insufficient electroweak 
phase transition.
Some extensions of the MSM resolve these problems and allow generation 
of the observed baryon asymmetry.
We can rule out some models such as the MSM and
give constraints to parameter spaces for other models, by requiring
successful baryogenesis and including other experimental results.
However, many extended models can explain the observed baryon asymmetry,
and we cannot specify the model beyond the MSM by these studies now.

Grand unified theories (GUTs), which unify the strong and electroweak 
interactions,
predict baryon number violation at the tree level.
The first baryogenesis scenario was proposed in the context of the GUT
\cite{Yoshimura}.
There, decays of heavy bosons cause departure from thermal equilibrium.
These scenarios can explain the observed baryon asymmetry,
although there remain some problems such as
super-heavy monopoles which overclose the universe, and
the washing out of the generated asymmetry by the sphaleron processes.

String theory is a promising candidate for the unified theory including gravity.
It has no symmetry assuring the baryon number conservation,
thus the baryon number is thought to be violated at the string scale or the 
Planck scale.
We pointed out a possibility of baryogenesis at the string scale 
\cite{Aoki Kawai}.
Even if the MSM describes the nature well above the electroweak scale,
it should be modified at the string scale or the Planck scale due to 
the gravitational effects.
Hence it is important to consider these scenarios.

Many other interesting mechanisms have been proposed.
In the supersymmetric extensions of the MSM, superpartners of the quarks and 
leptons are introduced and these scalars could have nonvanishing vacuum 
expectation values along the flat directions.
They oscillate afterwards, which generates 
baryon asymmetry \cite{AD}.
Topological defects such as cosmic strings and domain walls,
could affect the generation of the baryon asymmetry \cite{Bran}. 

In this article, we mainly consider two baryogenesis scenarios:
the electroweak baryogenesis and 
the string-scale baryogenesis.
Readers who are interested in the GUT-scale baryogenesis should
consult Ref. \cite{Kolb Turner}, for example.

In section \ref{sec:bau} we review the observational evidence 
for the existence of a baryon asymmetry of the universe,
and three necessary conditions to generate a baryon asymmetry dynamically.
In section \ref{sec:ewbg} we review the electroweak baryogenesis,
and in section \ref{sec:stringbg} we present  the string-scale baryogenesis.
The last section is devoted to conclusions and discussions.

\section{Baryon Asymmetric Universe}\label{sec:bau}
\setcounter{equation}{0}
\subsection{Evidence for a Baryon Asymmetry of the Universe}
Within the solar system we do not observe any bodies of antimatter.
High energy cosmic rays are generally believed to be of extra solar origin.
The small ratio of antimatter to matter in the cosmic rays, which is consistent
as secondary products, is an evidence for baryon asymmetry in our galaxy.
The existence of both matter and antimatter galaxies in a cluster of galaxy
would lead to a significant $\gamma$ ray flux from nucleon antinucleon 
annihilation.
The observed $\gamma$ ray flux suggests that clusters like Virgo cluster
must not contain both matter and antimatter galaxies.
Therefore we can say that there is  matter dominance over antimatter
on  scales at least as large as 
$(1-100)M_{\rm galaxy} \approx 10^{12}-10^{14}M_{\odot}$\footnote{
It is reported that the domain of matter dominance should be as large as
the entire visible universe (a few Gpc),
in order that the cosmic diffuse $\gamma$-ray 
spectrum near 1 MeV may not exceed the observational limits
\cite{CRG}. 
}.

We usually characterize the baryon asymmetry by the ratio, 
$n_B/s \equiv (n_b - n_{\bar{b}})/s$,
where $n_b$ is the number density for baryons, $n_{\bar{b}}$ is that of 
anti-baryons, and $s$ is the entropy density. 
This ratio remains constant even in the expanding universe,
when there are no baryon number violating processes
and no entropy productions.
The most stringent constraint on this ratio comes from the big bang 
nucleosynthesis.
The observed primordial abundances for light elements, ${\rm D}$, $^3 {\rm He}$,
$^4 {\rm He}$, and $^7{\rm Li}$, give
$\eta \equiv n_B/n_{\gamma} = (1.5-6.3)\times 10^{-10}$\cite{PDG}.
$n_{\gamma} = 2\zeta(3)/\pi^2 T^3 \approx 400 {\rm cm}^{-3}(T/2.7K)^3$
is the photon number density.
Since $s = 2\pi^2/45 g_{*s} T^3 \approx 7.04 n_{\gamma}$ at present,
\beq
n_{B}/s = (0.21-0.90) \times 10^{-10}.
\eeq
This is what we usually refer to as the baryon asymmetry of the universe.

\subsection{The Tragedy of a Symmetric Cosmology}

One might think that the whole universe is baryon symmetric but there are 
baryon-antibaryon fluctuations in space and we live in a body of matter.
However, as we show below, 
there is no plausible mechanism to separate matter from 
antimatter on such large scales as galaxy scales. 

\underline{Statistical Fluctuations}:
Although the matter-antimatter asymmetry appears to be large today 
in the sense $(n_{b}-n_{\bar b})/n_{b} \sim 1$,
the fact that $n_{B}/s$ is small implies that at very early times the asymmetry 
was small.
When the temperature of the universe was higher than $1 \GeV$,
baryons and antibaryons should have been about as abundant as photons, 
$n_{b} \approx n_{\bar b} \approx n_{\gamma}$.
Since $n_{B}/s$ is a conserved quantity and the entropy density is 
$s \approx g_{*}n_{\gamma}\approx 10^2 n_{\gamma}$ at $T \gsim 1 \GeV$,
\[
(n_{b}-n_{\bar b})/n_{b} \approx (n_{b}-n_{\bar b})/n_{\gamma}
                         \approx g_{*}n_{B}/s
                         \sim 10^{-8} (T\gsim1\GeV).
\]

This small asymmetry might be explained by the statistical (Poisson) 
fluctuations.
The co-moving volume that encompasses our galaxy today contains about
$10^{69}$ baryons and $10^{79}$ photons.
When $T\gsim1\GeV$ it contains about $10^{79}$ baryons and antibaryons.
From statistical fluctuations one can expect the asymmetry,  
$(n_{b}-n_{\bar b})/n_{b} \approx 1/\sqrt{n_{b}V} \sim 10^{-39.5}$,
which is too small.
Therefore it is impossible to explain the fact that we observe the baryon
asymmetry $n_{B}/s \sim 10^{-10}$ over the galaxy scales,
by statistical fluctuations.

\underline{Hypothetical Interactions}:
As the universe cooled down below $1 \GeV$
the number of baryons and antibaryons decreased tracking the equilibrium 
abundance, as long as annihilation processes were in equilibrium.
At $T \sim 20 {\rm MeV}$ the annihilation processes freeze out.
The residual ratio of baryon number density to entropy density is about
$n_{b}/s = n_{\bar b}/s \sim 10^{-19}$, which is too small.
In order to avoid this annihilation catastrophe, suppose that
some hypothetical interactions separated matter and antimatter 
before $T \simeq 38 {\rm MeV}$ when $n_{b}/s = n_{\bar b}/s \sim 10^{-10}$.
At that time, however, the causal region (horizon) was small and contained 
only about $10^{-7} M_{\odot}$.
Hence we cannot explain the asymmetry over the galaxy scales.

\subsection{Sakharov's Conditions}\label{sec:sakharov}
Since the baryon symmetric universe is hopeless, 
we must consider an excess of baryons over antibaryons in the whole 
universe\footnote{
One may consider a symmetric universe where baryogenesis and 
``anti-baryogenesis'' occur in different domains.
This is possible if we consider spontaneous $CP$ violation and 
inflation universe, for example.
}.
Three ingredients are necessary to dynamically generate 
baryon asymmetry 
from a baryon-symmetric initial state \cite{Sakharov}:\\
(1) baryon number nonconservation,\\
(2) violation of both $C$ and $CP$ invariance,\\
(3) departure from thermal equilibrium.

Without baryon number nonconserving interactions the universe remains 
baryon-symmetric and baryogenesis cannot occur.
Although we have not observed baryon number violating processes in laboratories 
yet, we believe that there are such interactions.
For example, in the electroweak theory baryon number conservation is 
violated due to chiral anomaly.
Grand unified theories also predict baryon number violating processes.
Also, at the Planck scale or the string scale baryon number 
is thought to be violated.

Under $C$ (charge conjugation) and $CP$ (charge conjugation combined 
with parity),
the $B$ (baryon number) of a state changes its sign.
Thus a state that is either $C$ or $CP$ invariant must have $B=0$.
If the universe begins with equal amounts of matter and antimatter, and without 
a preferred direction, then its initial state is both $C$ and $CP$ invariant.
Unless both $C$ and $CP$ are violated, the universe will remain either 
$C$ or $CP$ invariant as it evolves, and thus cannot develop a net baryon number
even if B is not conserved.
Hence, both $C$ and $CP$ violations are necessary.
$C$ is maximally violated in the weak interactions.
$CP$ violation has been observed in the $K$-on system.
However the fundamental understanding of $CP$ violation is still lacking,
and the studies of baryogenesis is expected to shed light on its understanding.

When baryon-number-changing processes are in chemical equilibrium,
the chemical potential for the baryon number vanishes.
$CPT$ invariance guarantees the mass equality of baryon and  antibaryon.
Thus the equilibrium distribution for baryons coincides with that 
for antibaryons, and thus $n_{b}=n_{\bar b}$.
Hence, in equilibrium with active baryon-number-changing processes, 
no baryon asymmetry is generated and 
any baryon asymmetry generated before is washed out.
However, in the context of the standard big bang cosmology,
the universe has experienced nonequilibrium many times,
when baryon asymmetry can be generated:
Phase transitions accompanied by symmetry breakings, 
freeze outs of some interactions due to the expansion of the universe,
$etc.$.

\section{Electroweak Baryogenesis}\label{sec:ewbg}

\subsection{Overview}\label{sec:ovewbg}

In the standard model of electroweak interactions,
the baryon number (and lepton number) are not conserved due to chiral anomaly
\cite{'t Hooft}.
The transition rates of these processes are small at zero temperature,
but enhanced to the observable order of magnitude at high temperatures
(subsection \ref{sec:bvew}).
This leads to the fact that any pre-generated baryon asymmetries may be
washed out
by these anomalous processes in thermal equilibrium,
which gives constraints on the models,
as we see in subsection \ref{sec:washout}. 
 
In subsection \ref{sec:mechanism}, we see how baryon asymmetry is generated
by the anomalous electroweak processes provided that
 the theory has $CP$ violation and
there is departure from thermal equilibrium.
The observed value of the baryon asymmetry can be obtained 
quite easily in these mechanisms.

However, within the minimal standard model (MSM), it is difficult to generate
the observed baryon asymmetry, as we see in subsection \ref{sec:bgmsm}.
First of all, $CP$ violation coming only from the CKM phase is too small to 
explain the observed value.
Secondly, the electroweak phase transition should be a strong 
first order phase transition, 
in order to realize the departure from thermal equilibrium,
and to preserve the generated baryon asymmetry.
In the MSM, however, these conditions are not satisfied.

In some extensions of the MSM, the above mentioned problems are resolved
and the observed baryon asymmetry can be generated at the electroweak scale.
The requirements of successful baryogenesis along with some experimental 
results rule out some models (such as the MSM),
and constrain the allowed parameter regions for other models.
In subsection \ref{sec:beyondmsm}, we see some extensions of the MSM
and their viabilities.

\subsection{Baryon Number Violation in the Electroweak Theory}\label{sec:bvew}

\subsubsection{Anomaly and Topological Structure of Vacua}

In the standard model of electroweak interactions,
the baryon number and lepton number are exactly conserved at the classical level.
However, due to chiral anomaly,
they are not conserved on the quantum level \cite{'t Hooft}: 
\beq
\partial_{\mu}j^{\mu}_{B} = \partial_{\mu}j^{\mu}_{L} 
= N_{f}[\frac{g^2}{16 \pi^{2}}\mbox{tr}(W^{\mu\nu}\tilde{W}_{\mu\nu})
        -\frac{g'^2}{32 \pi^{2}}B^{\mu\nu}\tilde{B}_{\mu\nu}],
\label{eq:anomaly}
\eeq
where $N_{f}$ is the number of generations, $W^{\mu\nu}$ and $B^{\mu\nu}$ are
the gauge field strength of $SU_L(2)$ and $U_Y(1)$, respectively, 
$g$ and $g'$ are their gauge couplings, and 
$\tilde{W}_{\mu\nu} = \frac{1}{2}\epsilon_{\mu\nu\alpha\beta}W^{\alpha\beta}$.
Note that the difference $B-L$ is strictly conserved.
The right hand side of this equation is a total divergence:
\beqan
\frac{g^2}{16 \pi^{2}}\mbox{tr}(W^{\mu\nu}\tilde{W}_{\mu\nu}) 
&=& \del_{\mu} K^{\mu},\n
\frac{g'^2}{32 \pi^{2}}B^{\mu\nu}\tilde{B}_{\mu\nu}
&=& \del_{\mu} k^{\mu},
\eeqan
where
\beqa
K_{\mu} &=& g^2\epsilon_{\mu\alpha\beta\gamma}\frac{1}{8\pi^{2}}
               \mbox{tr}(W^{\alpha}\partial^{\beta}W^{\gamma}
                 +\frac{2}{3}g W^{\alpha}W^{\beta}W^{\gamma}) \n
        &=& g^2 \epsilon_{\mu\alpha\beta\gamma}\frac{1}{8\pi^{2}}
               \mbox{tr}(\frac{1}{2}W^{\alpha\beta}W^{\gamma}
                 -\frac{1}{3}g W^{\alpha}W^{\beta}W^{\gamma}),\\
k^{\mu} &=& g'^2 \epsilon_{\mu\alpha\beta\gamma}\frac{1}{32\pi^{2}}
               B^{\alpha\beta}B^{\gamma}.
                                                    \label{eq:chern current}
\eeqa                      
Thus, integrating the anomaly equation (\ref{eq:anomaly}), we obtain
\beq
         \Delta B = \Delta L =  N_{f} \Delta (N_{\rm cs}-n_{\rm cs}),
\label{eq:delta}
\eeq
where $N_{\rm cs}$ and $n_{\rm cs}$ are Chern-Simons number,
\beqa
N_{\rm cs} & = & g^2\frac{1}{8\pi^{2}}
              \int d^{3}x \epsilon_{ijk} \mbox{tr}(W^{i} \partial^{j} W^{k}+
              \frac{2}{3} g W^{i}W^{j}W^{k}) \nonumber \\
           & = & \frac{1}{8\pi^{2}}
               \int d^{3}x \epsilon_{ijk} \mbox{tr}(\frac{1}{2}G^{ij}W^{k}
              -\frac{1}{3} g W^{i}W^{j}W^{k}),\\
n_{\rm cs} &=& g'^2 \frac{1}{32\pi^{2}}\int d^{3}x \epsilon_{ijk} B^{ij}B^{k}.
\label{eq:chern number}
\eeqa

\begin{figure}
\epsfxsize = 10 cm
\epsfysize =  8 cm
\centerline{\epsfbox{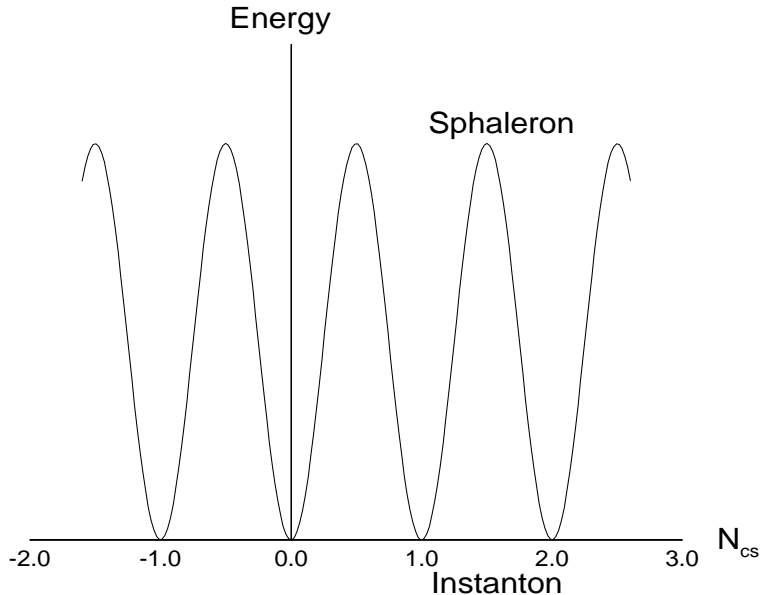}}
\caption{The structure of vacua of non-Abelian gauge theory.
         There are an infinite number of  
         degenerate vacua labeled by the Chern-Simons
         number.
         The sphaleron solution stays at the top of the potential barrier,
         and the instanton solution is used in analyzing 
          quantum tunneling.}
\label{fig:vac}
\end{figure}

Next, let us consider the vacuum configurations of the gauge field.
The Chern-Simons number for $U_{Y}(1)$, $n_{\rm cs}$, is strictly zero
since the value of the field strength, $B^{ij}$, takes zero in this case.
However, in the non-Abelian gauge theory, 
the vacuum configuration is a pure gauge, 
$W_{\mu}= g^{-1}U \del_{\mu}U^{\dagger}$,
and the Chern-Simons number becomes
\begin{eqnarray}
N_{\rm cs} &=& -\frac{1}{24\pi^{2}}\int d^{3}x 
           \epsilon_{ijk}\mbox{tr}(U\partial^{i}U^{\dagger}
                 U\partial^{j}U^{\dagger}U\partial^{k}U^{\dagger}) \nonumber \\
       &\equiv& \omega (U),
\end{eqnarray} 
where $\omega(U)$ is a topologically invariant quantity of $\Pi_{3}(G)\simeq Z$
when we identify spatial infinities.
This shows that there are an infinite number
of degenerate vacua which are classified by the Chern-Simons number.
They cannot be transformed continuously one another 
within the vacuum configurations.
However, for a general configuration other than vacua, the Chern-Simons number
takes a fractional value 
and we can transform continuously one vacuum into another 
{\it via} general configurations.
Fig. \ref{fig:vac} shows the vacuum structure of the non-Abelian gauge theory. 

When a transition from one vacuum to a topologically distinct vacuum
occurs, baryon number is changed, as you can see from \Eq{eq:delta}.
In the next subsections, we consider the rate for baryon number violating
processes by
estimating the transition rate from one vacuum to another.

In the gauge-Higgs system with nonzero vacuum expectation value of the 
Higgs field,
there exists a saddle-point solution 
at the top of the barrier between the neighboring vacua in Fig. \ref{fig:vac}.
Such a solution is called ``sphaleron'' solution \cite{Manton}.
It is an unstable solution and is easy to decay to the vacua.
This solution was constructed by considering a noncontractable loop 
in the configuration space.
In the minimal standard model, the energy of the sphaleron solution 
depends on the ratio of the parameters $\lambda/g$ and 
varies between $7.9 \TeV$ for $\lambda=0$ and $13.7 \TeV$ for $\lambda = \infty$,
where $\lambda$ is the Higgs self coupling constant. 
\beq
E_{\rm sp} \simeq M_{W}/\alpha_{W}
           \simeq 10 \TeV.
\eeq
The fact that the barrier hight is finite suggests that
the transition between vacua is likely to occur 
when the energy or the temperature of the system becomes large
\cite{Christ,KRS}.

\subsubsection{Transition Rate at Zero Temperature}

Since the hight of the potential barrier is finite, $E_{\rm sp} \sim 10 \TeV$,
it might be possible that we observe baryon number violating processes in 
future collider experiments.
Let us consider the processes
\beq
q+q \rightarrow 7\bar{q}+3\bar{l}+n_{W}W+n_H H,
\eeq
where 
$n_W$ and $n_{H}$ are the numbers of the produced gauge bosons and Higgs bosons,
respectively, and they take arbitrary numbers.
The cross sections for these processes can be obtained by calculating
the Green functions semi-classically about instanton-like configurations.
Instanton is a classical solution of Yang-Mills theory in four-dimensional 
Euclidean space-time, which connects the topologically distinct vacua
\cite{BPST}.
We will see below that the modification of the instanton solution 
is useful to estimate the anomalous cross sections in high energies.

The cross sections for vacuum-to-vacuum transitions 
are exponentially suppressed by the factor 
of $e^{-2S_{{\rm I}}}$ where $S_{{\rm I}}$ is the instanton action 
\cite{'t Hooft}. 
These processes had been therefore considered as 
hopelessly unobservable.
However, Ringwald \cite{Ringwald} and Espinosa \cite{Espinosa} showed 
that the cross section grows to an observable order of magnitude
at high energies $\sim E_{\rm sp}$
when O(1/$\alpha$) number of  gauge and Higgs bosons are involved 
in the final state\footnote{
Since there is no instanton solution as an exact classical solution 
in the $SU(2)$ gauge-Higgs system,
they used the so-called constraint instanton configurations \cite{Affleck}.
It is reported that calculations with valley instantons gives the similar result
\cite{Harano Sato}.}. 
Unfortunately, 
their results break the unitarity bound in these energies,
suggesting that their leading-order estimations are too naive and
higher-order corrections must be considered at least in these energies.
Since then, a lot of modifications to their calculations have been made 
(for reviews see Ref.\cite{Mattis,Rub Shap}).
Among them, the techniques of calculating the quantum corrections 
for the final states are developed.
Khoze and Ringwald 
showed an optimistic result suggesting that the 
anomalous cross 
section reaches an observable order
without breaking the unitarity bound
at the energies of the order of $E_{{\rm sp}}$ ($\sim$ 10TeV), 
by ``valley method'' \cite{Khoze Ringwald}.  
In this method, the total cross section in the instanton background 
is computed {\it via} optical theorem, 
by taking the imaginary part of the forward 
elastic scattering amplitude in the instanton-anti-instanton 
background,
so that the final state corrections are automatically included.

However, by considering the fermions in the intermediate states, 
we can see that
the cross sections for baryon number violating processes cannot grow so 
rapidly at high energies as their result \cite{Aoki1,Konishi3}.
Since we are interested in the ``anomalous'' cross sections,
we have to extract the correct part from the 
forward elastic scattering amplitude,
$i.e.$, the part where the fermion number in the intermediate state 
is topologically correct
in the background of instanton-anti-instanton configuration , and not 
just the same as that of the initial (and final) state.
It is known that when the incident energy is increased, 
the dominant configurations are the ones where the 
separation between instanton and anti-instanton is much smaller than the
instanton size \cite{Khoze Ringwald}.
Since such configurations seem to be
non-anomalous, the anomalous
part may be only a tiny fraction of the optimistic estimate of the 
cross section. 

Another severe suppression comes from 
the initial state corrections,
while in the valley method 
only the final state corrections are considered.
This can be understood intuitively as follows:
since the sphaleron solution has the spacial extension of the order of 
$M_{W}^{-1}$ \cite{Manton},
it is difficult to create such a configuration from the initial state of 
high energies $\sim E_{\rm sp}$.
On the other hand, when the initial state consists of a large number 
(O($1/\alpha_{W}$)) of particles whose energies are about $M_W$,
as in the case of high temperature systems,   
the transition can occur at unsuppressed rate. 
Indeed, the formulations were proposed, where the S-matrix element and the cross
section are calculated by functional integrations
\cite{KRT,GHP}.
The dominant classical configurations are given by solving the equation of 
motion with correct boundary conditions imposed by the initial and final states. 
In the case where many particles are involved in both initial and final 
states, the solutions exist and the cross sections are enhanced in high energies.
However, when the initial state involves a few particles,
solutions have not been discovered yet, suggesting that the transitions will be
suppressed even in high energies
(for review, see \cite{Rub Shap}).

\subsubsection{Transition Rate at Non-Zero Temperature}

As discussed in the previous subsections, the transition rate of the 
anomalous processes is expected to be enhanced at high temperatures.

Langer \cite{Langer} and Affleck \cite{Affleck} developed a technique for 
evaluating the transition rate of a system over a barrier at finite temperature.
They reduce the theory to one dimension, compute the flux of the 
system across the barrier in this direction weighted with a Boltzmann factor
for each possible initial state, and finally reintroduces the degrees of freedom 
transverse to this one mode.
The formula for the transition rate is
\beq
\Gamma \simeq \frac{\omega_{-}}{\pi T}{\rm Im}F,
\eeq
where $F$ is the free energy of the system evaluated about the configuration
at the top of the potential barrier (the sphaleron solution), 
and $\omega_{-}$ is the negative mode frequency for this configuration.
Note that the sphaleron solution is unstable and has the negative mode,
which gives the imaginary part of the free energy.
The transition rate per unit volume 
was estimated in the $SU(2)$ gauge-Higgs system
\cite{Arnold Mc, CLMW}:
\beq
\Gamma_{\rm sp} = \gamma (\alpha_{W} T)^{-3} M_{W}^7 e^{-E_{\rm sp}/T}\;\;
({\rm broken\; phase}),
\label{eq:spbro}
\eeq
where the constant $\gamma$ involves the Gaussian integrations for the 
sphaleron solution, and was estimated numerically in Ref. \cite{CLMW}.
We refer to this rate as the ``sphaleron transition rate''.
This rate is strongly suppressed by the Boltzmann factor up to temperatures
of hundreds of GeV.    

In the symmetric phase where the Higgs field has a vanishing vacuum expectation 
value, the above results are no longer valid.
We could find a path  which connects topologically distinct vacua 
in the configuration space, with arbitrary small potential energy at each point.
Although there is no reliable theory to estimate the transition rate,
a simple dimensional argument allows us to estimate it:
In order to generate an $O(1)$ change in baryon number 
(and Chern-Simons number), the anomaly equation (\ref{eq:anomaly}) implies
that a gauge field of spatial extent $R$ must have a field strength of order
$(gR^2)^{-1}$.
The energy of these configurations are of the order of $(g^2R)^{-1}$.
Thus, the smallest value of $R$ for which thermal transitions through such 
configurations are not strongly suppressed by the Boltzmann factor is 
$R \sim (g^2 T)^{-1}$.
This in fact is the dominant spatial scale, 
since the entropy favors the smaller configurations. 
From dimensional arguments,
the transition rate per unit volume will be proportional to $R^{-4}$,
and thus the result is\footnote{
There is an argument that damping effects in the plasma suppress the
sphaleron transition
rate by an extra factor of $\alpha_{W}$ \cite{ASY}.}
\beq
\Gamma_{\rm sp} = \kappa (\alpha_{W} T)^4\;\;({\rm symmetric \; phase}).
\label{eq:spsym}
\eeq 
Although there is no sphaleron solution in the symmetric phase,
we call ``sphaleron transition rate'' in this case, too.
The formula (\ref{eq:spsym}) was checked numerically, 
and the coefficient $\kappa$ was given to be of order one \cite{Ambjorn}.
The procedure is first generating an ensemble of configurations according to 
the Boltzmann weight, 
then considering the time evolution of  each configuration
by solving the equation of motion,
calculating the Chern-Simons number for it,
and finally taking the ensemble average over the initial configurations.
The result $<N_{cs}(t)^2>_T$ can be fitted to the expression for the random walk 
$\Gamma V t$, and the transition rate $\Gamma$ is obtained.

\subsection{Survival of Primordial Baryon Asymmetry}\label{sec:washout}

Before going into the stories of how baryon asymmetry is generated by 
the sphaleron processes, 
we consider how the primordial baryon asymmetry is (or is not) 
washed out by them.
Once the sphaleron transition is in equilibrium, any primordial
baryon asymmetry is washed out, 
as mentioned in subsection \ref{sec:sakharov}.
The sphaleron transition rate (\ref{eq:spbro}), (\ref{eq:spsym})
exceeds the expansion rate of the universe ($\sim g_{*}^{1/2}T^2/m_{\rm pl}$),
where $m_{\rm pl}$ is the Planck mass and $g_{*}$ is the number of 
relativistic species,
in the following region of temperatures:
\beqa
&&T_{\rm min} < T < T_{\rm max},\n
&&T_{\rm min} \sim T_{c} \sim 100 \GeV, \n
&&T_{\rm max} \simeq \alpha_{W}^4 m_{\rm pl} /g_{*}^{1/2} \simeq 10^{12} \GeV,
\eeqa
where $T_{c}$ is the critical temperature of the electroweak phase transition.

First, let us consider the case where 
baryon asymmetry was produced at temperature
$T > T_{\rm max}$, for example, the GUT scale or the string scale.
Since $B-L$ is conserved in the sphaleron processes,
one might think that $B+L$ is  washed out and 
survival baryon and lepton asymmetries are
$B = -L = \frac{1}{2}(B-L)_{\rm initial}$,
where subscript ``initial'' refers to the primordial asymmetry \cite{KRS}. 
In fact, the situation is more complicated:
The sphaleron processes only involve the left-handed fermions, and 
the charge neutrality of the universe must be preserved.
Thermodynamical calculations lead to the results
\beqa
B &=& \frac{8N_f+4m}{22N_f+13m}(B-L)_{\rm initial},\n
L &=& -\frac{14N_f+9m}{22N_f+13m}(B-L)_{\rm initial}
\;\;{\rm (symmetric\;phase)},
\eeqa
for the case in the symmetric phase, and
\beqa
B &=& \frac{8N_f+5(m+2)}{24N_f+13(m+2)}(B-L)_{\rm initial},\n
L &=& -\frac{16N_f +9(m+2)}{24N_f+13(m+2)}(B-L)_{\rm initial}
\;\;{\rm (broken\;phase)},
\eeqa
for the broken phase, 
where $N_f$ and $m$ are the number of generations and Higgs bosons, respectively
\cite{Harvey Turner}.
Anyway, if the primordial baryon asymmetry is generated by the 
$(B-L)$ conserving processes (which is the case in the minimal $SU(5)$ GUT
model \cite{GeoGla}), it is completely washed out.

Next, let us consider the case where there are some new interactions, 
which violate lepton number or baryon number, and are in thermal
equilibrium for some period between $T_{\rm min}$ and $T_{\rm max}$.
Combining them with the anomalous electroweak interactions,
both baryon and lepton asymmetries are washed out.
In order to avoid this problem, the new interactions must freeze out 
above the temperature $T_{\rm max}$.
For example, let us consider a lepton number violating interaction
\beq
{\cal L} = \frac{m_{\nu}}{v^2} l l \phi\phi,
\eeq
where $m_{\nu}$ is the neutrino mass, $v$ is the vacuum expectation value of the 
Higgs field, and $l$ and $\phi$ are the lepton and Higgs doublets, respectively.
The above mentioned condition for the survival of the primordial baryon asymmetry
gives an upper bound on the neutrino mass
\cite{FukYan,Harvey Turner}\footnote{
In fact, since the Yukawa coupling is small, the right-handed lepton number is 
effectively conserved. Thus this constraint is weakened \cite{CKO}.
}:
\beq
m_{\nu} \lsim g_{*}^{1/2}v^2/m_{\rm pl}\alpha_{w}^2 \sim 10^{-1}\eV.
\eeq  

Finally, let us consider the case where a baryon asymmetry is generated
by the sphaleron processes at the first order
electroweak phase transition.
Even if a baryon asymmetry is generated, 
it is washed out in the broken phase (true vacuum) just after generated,
if the sphaleron transition processes in the broken phase are in equilibrium.
The sphaleron transition rate in the broken phase at 
the electroweak phase transition,
depends sensitively on the sphaleron energy (see \Eq{eq:spbro}),
which will be proportional to the vacuum expectation value of the 
Higgs field.
Thus the condition for the survival of the generated baryon asymmetry is 
\cite{Shap}
\beq
v(T_{c})/T_{c} \gsim 1,
\label{eq:washout}
\eeq
where $v(T_{c})$ is the vacuum expectation value of the Higgs field 
in the broken phase at the critical temperature, $T_c$.
 
\subsection{Mechanisms of Electroweak Baryogenesis}\label{sec:mechanism}

Since the baryon number violating processes occur rapidly at high temperatures,
a baryon asymmetry is generated 
if there is a bias to prefer the direction of increasing baryon number.
From detailed balance, we obtain the equation
\beq
\frac{dn_{B}}{dt} = -N_{f}^2\frac{\mu_{B}\Gamma_{\rm sp}}{T},
\label{eq:master}
\eeq
where  $N_{f}$ is the number of flavors,
$\Gamma_{\rm sp}$ is the sphaleron transition rate (\ref{eq:spbro})
(\ref{eq:spsym}), and $\mu_{B}$ is the chemical potential 
for baryon number\footnote{
Another form of this equation is written in Ref. \cite{JPT}:
$\frac{dn_{B}}{dt} = -N_{f}n_{f_L}\frac{\Gamma_{\rm sp}}{T^3/6}$,
where $n_{f_L}$ is 
the number density of left-handed fermions.}.
The factor $N_{f}^2$ arises since $N_f$ baryon number is changed
in one sphaleron transition.
From dimensional analysis, we can expect that the baryon-to-entropy ratio
of the order of
\beq
\frac{n_{B}}{s} \sim N_{f}^2\frac{\kappa \alpha_{W}^4}{g_{*}} \sim 10^{-7}
\label{eq:expected value}
\eeq
will be generated, provided that
there is enough nonequilibrium distributions of matter,
and the theory has enough $CP$ violation.

As we will see below, if the electroweak phase transition is first order,
nonequilibrium distribution is realized around the bubble wall.
If we consider the two-Higgs-doublet model, for example,
there is enough $CP$ violating phase in the Higgs potential.
The vacuum expectation values of the Higgs fields take complex values in general,
and their relative phase cannot be reduced to zero by field redefinition.
The bubble wall at the phase transition will have a profile like
\beq
\phi(z) = v[\frac{1-\tanh(z/l_w)}{2}]
          e^{-i\Delta\theta[1+\tanh(z/l_w)]/2},
\label{eq:wall profile}
\eeq
where $z$ is the coordinate perpendicular to the wall, $l_w$ is the wall width,
and $\Delta\theta$ is the difference of the complex phases 
between in the symmetric phase and in the broken phase.
Through the Yukawa couplings, fermions interact with this 
$CP$ violating bubble wall, which can cause nonzero chemical potential for
baryon number.

Let us consider characteristic time scales governing the relevant reactions:
The time scale for the universe expansion, $\tau_{H}$,
the sphaleron transition, $\tau_{\rm sp}$,
the thermalization processes due to strong and weak interactions, $\tau_{\rm th}$,
and the time scale to cause nonequilibrium distributions, 
measured by how fast the wall passes through 
a point in the plasma, $\tau_{\rm wall}$.
At the electroweak phase transition, they are estimated to be
\cite{DLHLL}
\beqa
\tau_{H}^{-1} &\simeq& 1.66 g_*^{1/2} \frac{T^2}{m_{\rm pl}} \sim 10^{-16}\: T,\n
\tau_{\rm sp}^{-1} &\simeq& \alpha_{W}^4 T \sim 10^{-6}\: T, \n
\tau_{\rm th}^{-1} &\sim& \left\{ \begin{array}{l}
                     0.25 \:T \;\; {\rm for \;quarks}\\
                     0.08 \:T \;\; {\rm for \;leptons},
                     \end{array} \right.\n
\tau_{\rm wall}^{-1} &\simeq& \frac{v_w}{l_w} \sim (0.01 - 1) \:T,
\eeqa
where $v_w$ is the velocity of the wall.
Since $\tau_H^{-1} \ll \tau_{\rm sp}^{-1} \ll \tau_{\rm wall}^{-1}$,
the sphaleron transitions cannot be out of equilibrium by the expansion of the 
universe,
but they can be out of equilibrium in the first order 
electroweak phase transition.

Let us consider the following two extreme cases:\\
1) $\tau_{\rm th}^{-1} \ll \tau_{\rm wall}^{-1}$, thin wall case.
In this case, the fermions interact with the bubble wall 
like quantum mechanical particles, reflected and transmitted
by a potential barrier,
and their interactions with the particles in the plasma are neglected,
when they propagate across the wall. \\
2) $\tau_{\rm th}^{-1} \gg \tau_{\rm wall}^{-1}$, thick wall case.
In this case, the fermions are scattered by the particles in the plasma,
which can be described by the thermodynamics in equilibrium,
and the quantum reflection by the wall can be neglected.\\
Mechanisms for baryogenesis were proposed \cite{CKN1, CKN2}
in these two cases.
In both cases, the expected value (\ref{eq:expected value}) is obtained
with  the optimal parameters for $\Delta\theta$, $l_{w}$, {\it etc.},
thus the observed baryon asymmetry can be explained in quite
large regions of parameter space.

\subsubsection{The Thin Wall Case: Charge Transport Mechanism }

Since the bubble wall has a $CP$ violating complex phase 
(\ref{eq:wall profile}), 
when the fermions are scattered off the wall,
a net charge is reflected from the wall,
which is transported into the region preceding the bubble.
This charge asymmetry can bias the sphaleron transitions in the symmetric phase
and generate a baryon asymmetry. \cite{CKN,CKN2}

The procedure to calculate the baryon asymmetry in this mechanism is 
as follows:\\
0) Compute the wall profile and velocity, including the space dependent 
$CP$ violating phase, from the finite temperature effective potential
(or effective action).
The form of $CP$ violating bubble wall is studied \cite{FKOT}, 
although in Ref. \cite{CKN2}, the wall profile is assumed to be  
(\ref{eq:wall profile}).\\
1) Calculate the reflection coefficients for fermions striking the wall,
by integrating the Dirac (or Majorana) equation of motion 
in the wall-rest frame.
Because of $CP$ violation, the reflection coefficient for particles and 
antiparticles can be different.
\beq
\Delta R \equiv R_{R \rightarrow L}-R_{\bar{R} \rightarrow \bar{L}}
         \neq 0
\eeq
2) Convolute the reflection coefficients with the incoming (boosted) flux of 
particles in the plasma, to determine the outgoing flux from the wall 
into the symmetric phase.
Due to the nonzero
difference between the reflection coefficients for particles and 
antiparticles,
this flux carries nonzero quantum numbers $X$;
for example, chiral charge.
In Ref. \cite{CKN2}, they consider the hypercharge since it is conserved
in the symmetric phase.\\
3) The nonzero flux from the wall, $F_X$ leads to a nonzero
charge density distribution, $n_X(z)$ in the region preceding the wall.
One can calculate how far the charge is transported into the symmetric phase,
by Monte Carlo simulation or solving the Boltzmann equation.\\
4) The nonzero charge density  causes a nonzero chemical potential
for baryon number:
\beq
\mu_B = C \frac{n_X}{T^2},
\eeq
where $C$ is a constant of the order one,
and is given by thermodynamical calculations.\\
5) One can calculate the baryon number density,
by integrating the equation (\ref{eq:master})
\beqa
n_B &=& -\int dt N_f^2\frac{\mu_B \Gamma_{\rm sp}}{T} \n
    &\simeq& -C N_f^2\frac{\kappa \alpha_W^4 T}{v_w} \int_0^{\infty} dz\: n_X(z),
\eeqa
where we assume that the sphaleron transition rate in the broken phase is 
negligible.
Thus, in the above equation,
the charge density is integrated from the wall to the symmetric phase.
While the bubble wall sweeps the whole universe,
the nonzero baryon number density is generated in any point of the space. 

The baryon-to-entropy ratio is 
\beq
\frac{n_B}{s} \simeq -C N_f^2 \frac{\kappa \alpha_{W}^4}{g_*}
                   \frac{F_{X}}{v_w T^3} \tau T,
\eeq
where $\tau$ is the transport time within which the scattered fermions are
transported into the symmetric phase and captured by the wall. 
The result of Monte Carlo simulation shows $\tau T \sim 10-10^3$ for the top 
quark \cite{CKN2},
rather large value compared with its diffusion constant $D \sim T^{-1}$.
It is because particles are scattered primarily in forward direction. 
The baryon-to-entropy ratio reaches $10^{-7}$,
when the top quark scattering is considered, the $CP$ violation parameter
$\Delta\theta$ is taken to be of the order one, and the wall width 
is assumed to be of the order $T^{-1}$ \cite{CKN2,FKOT2}.  

\subsubsection{The Thick Wall Case: Spontaneous Baryogenesis }

In the thick wall case, one can treat the plasma around the bubble wall
as being in quasi-equilibrium, with a classical time-dependent field.
However, some reactions such as baryon number violating processes are 
not in chemical equilibrium,
since $\tau_{\rm sp}^{-1} \ll \tau_{\rm wall}^{-1}$.
Thus the $CP$ violating complex phase in the bubble wall (\ref{eq:wall profile})
may give nonzero chemical potential for baryon number. \cite{CKN,CKN1}

Too see this, redefine the fermion fields by phase factor, to
remove the phase from the Yukawa interaction.
This leads to a new interaction from the fermion kinetic energy term:
\beq
{\cal L_{\rm K.E.}} \rightarrow {\cal L_{\rm K.E.}} + \del_{\mu}\theta j^{\mu},
\eeq
where $j^{\mu}$ is the current associated with the fermion-field redefinition.
From the time component of this new interaction,
we can see that $\dot{\theta}$ corresponds to the chemical potential for 
the charge associated with the current $j^{\mu}$.
In Ref. \cite{CKN1}, they redefine the fermion fields according to 
their hypercharge,
since the hypercharge is anomaly free so that generation of 
new interaction $\theta F \tilde{F}$ is avoided.

This nonzero ``chemical potential'' $\dot{\theta}$ causes a nonzero 
chemical potential for baryon number:
\beq
\mu_{B} = C' \dot{\theta},
\eeq 
where $C'$ is a constant of order one,
and is given by thermodynamical calculations. 
This bias generate a nonzero baryon number density,
which is calculated by integrating the \Eq{eq:master}:
\beqa
n_{B} &=& -\int dt N_f^2\frac{\mu_{B} \Gamma_{\rm sp}}{T} \n
      &\simeq& -C'N_f^2\kappa \alpha_{W}^4 T^3 (\Delta\theta)_{\rm co}.
\eeqa
Here, we assume that the sphaleron transition rate goes rapidly to zero
as the vacuum expectation value of the Higgs field approaches a value
$\phi_{\rm co} e^{i\theta_{\rm co}}$ in the bubble wall.
$(\Delta\theta)_{\rm co}$ is the difference between $\theta_{\rm co}$
and the value of $\theta$ in the symmetric phase.
Thus, the baryon-to-entropy ratio is 
\beq
\frac{n_{B}}{s}\simeq 
-C' N_f^2\frac{\kappa \alpha_{W}^4 }{g_*}(\Delta\theta)_{\rm co}.
\label{eq:ba sb}
\eeq

It was pointed out that when $\phi_{\rm co}$ takes a small value,
another small factor decreases the result (\ref{eq:ba sb})
and this scenario cannot generate enough amount of
baryon asymmetry \cite{Dine Thomas}.
In the symmetric phase,
where the vacuum expectation value of the Higgs field vanishes, 
its complex phase $\theta$ has no physical meaning,
and no baryon asymmetry is generated by this mechanism.
Accordingly, when the vev is small, there must be some suppression factor
to the result (\ref{eq:ba sb}).

Until now we consider the case where a baryon asymmetry is generated only in a 
small region within the bubble wall:
$z_{\rm co} < z < z_{\rm sym}$,
where $z_{\rm co}$ is the position in the wall where the sphaleron 
transition rate is cutoff, and $z_{\rm sym}$ is the
surface between the wall and the symmetric phase.
However, if we consider the effect of diffusion,
nonzero charges are diffused from within the bubble wall, and 
a baryon asymmetry can be generated in a wide region in the symmetric phase
\cite{CKN3,JPT}.
Solving the diffusion equation, one concludes that
the resultant baryon-to-entropy ratio is of the order of $10^{-7}$,
irrespective of the value of $z_{\rm co}$, the profile of the wall, and 
wall velocity.
Now it is established that transport of $CP$ violating quantum numbers
into the symmetric phase plays an important role in 
electroweak baryogenesis for 
all values of bubble wall widths.

\subsection{Baryogenesis within the Minimal Standard Model}\label{sec:bgmsm}

Within the MSM, it is difficult to generate
the observed baryon asymmetry of the universe.
First of all, $CP$ violation coming only from the CKM phase is too small to 
explain the observed value.
Secondly, the electroweak phase transition should be a first order phase 
transition in order to realize the departure from thermal equilibrium,
as we saw in subsection \ref{sec:mechanism}.
Moreover, in order to avoid the washing out of the generated baryon asymmetry,
the vacuum expectation value of the Higgs field in the
broken phase at the critical temperature must be sufficiently large
(subsection \ref{sec:washout}).
In the MSM, however, these conditions are not satisfied.

\subsubsection{$CP$ violation}

Within the MSM with three generations, a $CP$ violating complex phase is in 
the Cabibbo-Kobayashi-Maskawa (CKM) matrix \cite{KM}.
This phase could be rotated away if any pair of quarks
of the same charge were degenerate in mass,
or if any of the mixing angles vanished.
Thus, any physical observables related to $CP$ violation
will be proportional to a basis-invariant combination \cite{Jarlskog}
\beqa
d_{CP} &=& \sin(\theta_{12}) \sin(\theta_{23}) 
           \sin(\theta_{31}) \sin(\delta_{CP}) \n
       & & \times (m_t^2-m_c^2)(m_t^2-m_u^2)(m_c^2-m_u^2)
                  (m_b^2-m_s^2)(m_b^2-m_d^2)(m_s^2-m_d^2)).
\eeqa
The natural mass scale in the electroweak baryogenesis seems to be the
temperature, $T \sim 100\GeV$.
From dimensional arguments, $CP$ violating observables
will be suffer from a suppression factor
\beq
\frac{d_{CP}}{T^{12}} \sim 10^{-18}.
\eeq
Hence the baryon-to-entropy ratio will be of the order of
\beq
\frac{n_B}{s} \sim  N_f^2 \frac{\kappa\alpha_W^4}{g_*}\frac{d_{CP}}{T^{12}}
              \sim 10^{-25}.
\label{eq:ba dim}
\eeq

The above arguments would be invalid and the observed baryon-to-entropy ratio
could be generated,
if the natural scale were not the temperature but, for example,
the mass difference between the strange and down quarks. In Ref. \cite{FS},
they considered the scattering of the quarks off the bubble wall in the plasma,
and pointed out that in some energy regions,
the $s$ quark is totally reflected but the $d$ quark is partly 
reflected and partly transmitted, which enables us to avoid GIM cancellation.
However, the quasi-particles in the plasma decay rapidly due to the strong 
interactions,
and the resultant baryon asymmetry is not enhanced appreciably,
remaining at the value of \Eq{eq:ba dim} \cite{Huet Sather}. 

\subsubsection{Phase Transition}

The standard approach to analyzing phase transition is computing the 
finite temperature effective potential \cite{Dolan Jackiw}.
For high temperatures, the one-loop effective potential is approximated by
the high-temperature expansion:
\beq
V(\phi ,T) = D(T^2-T_0^2)-ET\phi^3+\frac{\lambda_T}{4}\phi^4,
\eeq
where 
\beqa
D &=& \frac{1}{8v_0^2}(2m_W^2+m_Z^2+m_t^2), \n
E &=& \frac{1}{4\pi v_0^3}(2m_W^3+m_Z^3) \sim 10^{-2}, \n
T_0^2 &=& \frac{1}{4D}(m_H^2-8Bv_0^2), \n
B &=& \frac{3}{64\pi^2 v_0^4}(2m_W^4+m_Z^4-4m_t^4), \n
\lambda_T &=& \lambda - \frac{3}{16\pi^2 v_0^4}
             (2m_W^4 \ln\frac{m_W^2}{a_BT^2} + m_Z^4 \ln \frac{m_Z^2}{a_BT^2}
              -4m_t^4 \ln \frac{m_t^2}{a_FT^2}).
\eeqa
Here $v_0 = 246 \GeV$ is the value of the Higgs field at the minimum of 
$V(\phi ,0)$, $\lambda = m_H^2/2v_0^2$ is the Higgs self coupling constant,
and $\ln a_B = 2 \ln 4\pi -2\gamma \simeq 3.51$,
$\ln a_F = 2 \ln \pi -2 \gamma \simeq 1.14$.

\begin{figure}
\epsfxsize = 10 cm 
\epsfysize = 8 cm
\centerline{\epsfbox{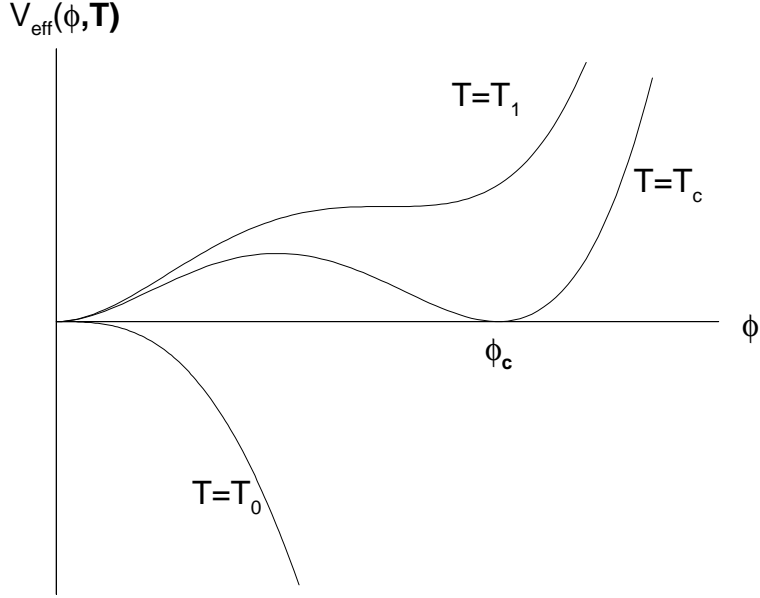}}
\caption{Qualitative form of the effective potential as a function of 
temperature. It seems to suggest a first order phase transition.
}
\label{fig:eff}
\end{figure}

The behavior of the effective potential is depicted in Fig. \ref{fig:eff}.
At very high temperature the only minimum of the potential is at $\phi = 0$.
A second minimum appears at $T = T_1$, where,
\beq
T_1^2 = \frac{T_0^2}{1-9E^2/8\lambda_{T_1} D}.
\eeq
The value of the field in this minimum at $T=T_1$ is equal to 
\beq
\phi_1 = \frac{3ET}{2\lambda_{T_1}}.
\eeq
The values of the potential in the two minima become equal to each other at
$T = T_c$, where,
\beq
T_c^2 = \frac{T_0^2}{1-E^2/\lambda_{T_c} D}.
\eeq
At that moment the field $\phi$ in the second minimum becomes equal to 
\beq
\phi_c = \frac{2ET_c}{\lambda_{T_c}}.
\label{eq:phic}
\eeq
The minimum of the potential at $\phi = 0$ disappears at $T=T_0$,
when the value of the scalar field in the second minimum becomes equal to 
$\phi_0 = 3ET_0/\lambda_{T_0}$.

It seems that 
at some lower temperature than $T_c$, first-order phase transition occurs.
When the Higgs mass is light and $\lambda$ is small,
$\phi_c$ and the potential barrier between the two minima become
large, suggesting the strong first-order phase transition.
Indeed, it was checked that if the Higgs boson mass is smaller than about
the masses of $W$ and $Z$ bosons,
bubbles of the new phase are nucleated
and expand to fill the whole universe \cite{DLHLL}.
The propagations of the bubble walls make the distribution of particles 
around the wall in nonequilibrium, 
which may generate a baryon asymmetry of the universe.  

In order to avoid the washing out of the generated baryon asymmetry,
the sphaleron transition rate must be small in the broken phase at the
phase transition, leading to the condition $\phi_c/T_c \gsim 1$ 
(\ref{eq:washout}).
Using \Eq{eq:phic} as a first approximation,
this condition gives an upper bound on $\lambda$,
and thus on the Higgs boson mass \cite{Shap}:
\beq
m_H \lsim 45\GeV
\eeq
It is already inconsistent with the present experimental lower bound 
$m_H \gsim 70\GeV$.
Hence, the observed baryon asymmetry cannot be explained within the MSM.

Next, let us consider higher-order corrections to the one-loop effective
potential presented above, and a reliability of perturbation expansions.
Theories with massless or light particles suffer from the problem of 
infrared divergences at high temperatures.
For example, the longitudinal component of gauge boson is screened in a plasma,
and the corresponding Debye mass is $m_{\rm Debye}^2 =11/6g_W^2T^2$
in one-loop approximation.
It becomes larger than the tree level contribution to the $W$ boson mass,
$m_W^2 = 1/4g_W^2\phi^2$ in the region $\phi/T < O(1)$,
suggesting that the naive perturbation expansion is not reliable.
This one-loop Debye screening effects can be resummed to all orders in 
perturbation expansions by substituting $m_W^2+m_{\rm Debye}^2$ for $m_W^2$ 
in the propagators.
This corresponds to the resummation of ring- or daisy-diagrams.   

The ring-resummed one-loop calculation indicates that 
the phase transition becomes weaker than in the naive one-loop calculation,
and the upper bound on the Higgs boson mass becomes more severe:
$m_H \lsim35 \GeV$ \cite{DLHLL}.
The ring-resummed two-loop calculation
indicates stronger phase transition and relaxes the upper bound:
$m_H \lsim 40 \GeV$ \cite{ArnEsp, FodHeb}.
In fact, since the top quark is heavy,
the phase transition becomes much weaker when its effect is included,
and the condition $\phi_c/T_c \gsim 1$
cannot be satisfied in any value of the Higgs boson mass. 

When the Higgs mass is large, the two-loop-order corrections become huge 
and the perturbation expansions do not converge even with the ring-resummation.
We should say that 
the perturbative calculations of the effective potential are not reliable
in the realistic range of the Higgs boson mass, $m_H \gsim 70 \GeV$,
and we should use another method to analyze electroweak phase transition
in the MSM. 

Lattice simulation is a non-perturbative method for the analysis of the
phase transition.
Two approaches have been applied for this study.
One is to employ the original four-dimensional model
\cite{BIKS,FHJJMC,YAoki}.
The other is to treat a three-dimensional model derived from the original model
by dimensional reduction in time direction
\cite{KLRS,IKPS}.
They analyzed the Monte-Carlo time history, the finite-size scaling,
the correlation length, the latent heat, the interface tension, {\it etc}.
When the Higgs mass is small, their results reproduce those of the 
perturbative analysis: the phase transition is of the first order,
and the strength of the phase transition becomes weaker as the Higgs boson
mass increases.
Moreover, they showed that the first-order phase transition terminates 
at around $m_H \sim 80\GeV$, and two phases are connected at larger Higgs masses.
Thus, if the Higgs mass is larger than this critical value,
no baryon asymmetry can be generated at the electroweak scale.

\subsection{Beyond the Minimal Standard Model}\label{sec:beyondmsm}

In contrast to the MSM, in most extensions of the MSM
there can be new $CP$ violation sources other than the CKM phase,
which could be an origin of the observed baryon asymmetry.
Also, since there are more parameters, 
the electroweak phase transition can be sufficiently
strong first-order to avoid the washing out of the generated baryon 
asymmetry, in some region of the parameter space.
However, other experimental constraints such as electric dipole moments (EDM) of 
neutron and atoms must be considered to check whether the model is viable.

In the two-Higgs-doublet model, there is a new $CP$ source in the Higgs 
potential and an enough amount of baryon asymmetry can be generated,
as we saw in subsection \ref{sec:mechanism}.   
The experimental constraints of EDM allow the $CP$ phase to be as large as O(1)
\cite{Barr}.
The experimental lower bound on the mass of the lightest Higgs boson and
the requirement of the sufficiently strong first-order phase transition 
can be simultaneously satisfied in a wide range of the parameter space
\cite{Tur Zad}.
Therefore, this model will be viable.

In the minimal supersymmetric standard model (MSSM)\footnote
{By MSSM we mean the supersymmetric extension of the standard model 
with minimal particle content and $R$ parity conservation, whereas
the constraint MSSM includes the assumption of coupling constant unification,
universal squark and gaugino masses and universal trilinear couplings 
at the unification scale.},
the phase transition can be sufficiently strongly first order 
only in a restricted region of the parameter space\footnote{
Two-loop calculations show stronger phase transition,
thus these bounds will be relaxed \cite{Esp}.}
\cite{CQW,DGFW}:
\beq
m_{\tilde{t_R}} \lsim 175 \GeV, m_h \lsim 80 \GeV, m_A \gsim200\ GeV,
\tan \beta \lsim 2.5, \tilde{A}_t \simeq 0.
\eeq
Here $m_{\tilde{t_R}}$ is the mass of the stop (supersymmetric 
partner of the top quark),
$m_h$ is that of the lightest Higgs boson,  
$m_A$ is that of the pseudoscalar Higgs boson,
$\tan \beta$ is the ratio of the vacuum expectation values of the two Higgs
bosons,
and $\tilde{A}_t =  A_t + \mu/\tan \beta$ is the effective 
$\tilde{t}_L-\tilde{t}_R$ mixing parameter.
The upper bound on the lightest Higgs mass, $80\GeV$, can be reached by the 
LEP2 experiments,
and the upper bound on the stop mass by the Tevatron.

There can be new $CP$ violating phases in the soft supersymmetry breaking terms
in the MSSM.
The observed value of the baryon asymmetry can be generated at the electroweak 
phase transition
if the new $CP$ phases are larger than about $10^{-2}$
\cite{HN,AOS,CQRVW}.
However the presence of these new phases also lead to the neutron EDM.
The experimental bound tells us that either these phases are less than about
$10^{-2}$ \cite{FPT} or the squark masses are very large $m_{\tilde{u}}\gsim 1\TeV$
\cite{KO}.        
Thus, if the present baryon asymmetry is generated by this mechanism,
either neutron EDM will be discovered soon or the first generation squarks 
are heavy.

Also, large mixings between the right-handed up-type squarks, 
$\tilde{c}_R-\tilde{t}_R$ or $\tilde{u}_R-\tilde{t}_R$,
along with the $CP$ violating phase already present in the quark mass matrix,
lead to the generation of the observed baryon asymmetry 
and the sufficiently strong electroweak phase transition
\cite{Worah}.
This scenario predicts $D-\bar{D}$ mixing at a level 
that should be discovered soon.
Thus the possibility of baryogenesis in the MSSM significantly constraints
its parameter space, and
near future experiments will shed light on this picture.

Other extended models such as a model with heavy neutrinos \cite{YS}
and a model with vector-like quarks \cite{USY,McD},
were studied and the observed baryon asymmetry can be generated  
in certain parameter regions in these models.


\section{String-Scale Baryogenesis}\label{sec:stringbg}

\subsection{Overview}
In this section we
consider baryogenesis scenarios at the string scale
or the Planck scale, and show how the observed baryon asymmetry can
be explained in these new scenarios
\cite{Aoki Kawai}.
Even if the minimal standard model (MSM) describes the nature well above the 
electroweak scale,
it must be modified around the string scale or the Planck scale
due to gravitational effects.
Hence it is important to consider the baryogenesis scenarios at these scales. 

At the string and Planck scales, the Sakharov's three necessary conditions for 
baryogenesis are satisfied.
In string theory, there is no symmetry assuring the baryon number conservation.
Also, when we consider string inspired models or 
effective theories with a cut-off at the string scale,
there is no reason to prohibit baryon- or lepton-number 
violating interactions in theories with a cut-off.
On the other hand, the MSM, 
which is required to be renormalizable and gauge invariant,
does not allow such interactions.
Sources of $CP$ violation at the string scale can differ 
from those at the electroweak scale,
and other sources than the CKM phase are allowed at the string scale.

As for departure from thermal equilibrium, we use
the so called Hagedorn temperature \cite{Hagedorn,Hagedorn2}.
String theory has a limiting temperature, where the higher excited states of 
string theory are occupied.
The decay processes of these states will cause nonequilibrium distributions.
Baryon asymmetry is generated during the decay of the higher excited states.
It is also generated after the decay 
since nonequilibrium distributions caused by the decay processes
are maintained 
until the rates for thermalization processes dominate the expansion rate
of the universe. 

The resultant baryon to entropy ratio will not have suppression factors
since the theory has only one scale, the string scale or the Planck scale.
Hence, we expect the observed value is obtained in these scenarios.

In subsection \ref{sec:model} we present a model and show
how it satisfies the three conditions for baryogenesis.
In subsection \ref{sec:Boltzmann} we calculate the resultant lepton asymmetry
by considering Boltzmann equations 
and show that these scenarios can explain the observed baryon asymmetry.
In subsection \ref{sec:comment} we give some comments.

\subsection{A model}
\label{sec:model}
In this subsection we present a model of string-scale baryogenesis. 
Here we consider the following case:
the nature is described by the MSM below the string scale.
There is not SUSY, nor GUT, nor inflation.
Indeed, non-SUSY, non-GUT string models have been proposed \cite{Dienes}.

Hence, as an effective theory of string theory, 
we consider a model whose matter content is the same 
as that of the MSM.
For simplicity, we consider lower-dimensional operators.
Let us consider the model,
\beqa
{\cal L} &=& {\cal L}_{\rm MSM} \n 
         & &  +\frac{1}{4} \frac{g_{st}^2}{m_{\rm st}} h_{ij} \epsilon^{\alpha \beta}
               (\epsilon_{ab} \epsilon_{cd}+\epsilon_{ad} \epsilon_{cb})
            l_{\alpha}^{ia} \phi^b l_{\beta}^{jc} \phi^d
           +{\rm h.c.}\n 
         & &  +\frac{1}{4} \frac{g_{st}^2}{m_{\rm st}^2} \epsilon^{\alpha \beta}
            \epsilon^{\gamma \delta}
            [C_{ijkl}(\delta_{ac} \delta_{bd} + \delta_{ad} \delta_{bc})
             +C'_{ijkl}(\delta_{ac} \delta_{bd} - \delta_{ad} \delta_{bc}) 
             +C''_{ijkl} \epsilon_{ab} \epsilon_{cd}] \n 
 & & \times l_{\alpha}^{ia} l_{\beta}^{jb} l_{\gamma}^{kc*} l_{\delta}^{ld*},
\label{eq:model}
\eeqa
where $l$'s are the lepton doublets and $\phi$ is the Higgs doublet.
The string coupling constant and the string scale are $g_{\rm st}$ and $m_{\rm st}$,
respectively. 
Also, $i,j,k$ and $l$ are generational indices,
$\alpha,\beta,\gamma$ and $\delta$ are spinorial indices, 
and $a,b,c$ and $d$ are $SU(2)_L$ gauge-group indices.

The first term violates lepton number conservation since it is a Majorana-type
coupling.
The second term is included to incorporate {\it CP} violation.
Through a unitary transformation,
the coefficient of the first term, $h_{ij}$, can be rotated into a real
diagonal form.
However, if the second term is present we cannot guarantee
that both terms can be rotated to real forms simultaneously.  
With two or more generations, this model violates {\it CP} invariance.
Henceforth, we consider exactly two generations for simplicity .

In the remainder of this subsection we speculate on the 
departure from thermal equilibrium.
First of all, let us consider how the universe would have been 
in the context of string theory if thermal equilibrium had been maintained
\cite{Hagedorn,Hagedorn2}.
Since the density of states is exponentially increasing in string theory,
it has a limiting temperature, the Hagedorn temperature,
\beq
T_{\rm H}=\left\{ \begin{array}{ll} 
                 m_{\rm st}/2 \sqrt{2}\pi \sim 5.93\times 10^{16}\:{\rm GeV}, 
               &({\rm type \;II})\\
                 m_{\rm st}/(2+\sqrt{2})\pi \sim 4.92\times 10^{16}\:{\rm GeV}.
               &({\rm heterotic \; string})
              \end{array}       
       \right.
\eeq
While the universe is contracting, the energy density increases 
but the temperature remains just below the Hagedorn temperature.
When the energy density is low compared to the string scale,
matter is  composed of particles, or the excitations of short strings  
whose lengths are of the order of $m_{\rm st}^{-1}$.
However, the high-energy limit of 
the single-string density of states is found to be \cite{Hagedorn2}
\beq
\omega(\epsilon) = \frac{\exp(\epsilon / T_{\rm H})}{\epsilon}.
\eeq
When we consider the microcanonical ensemble,
it turns out that long strings traverse the entire volume of the universe
in a sufficiently high energy density.

Next we consider whether thermal equilibrium is actually realized,
by comparing the rates for the thermalization processes with  the
expansion rate of the universe, the Hubble rate.
When matter is composed of particles, 
the rates for thermalization processes are
$\Gamma_{{\rm th}} \sim g_{*}\alpha_{\rm st}^{2}T$, and
the Hubble expansion rate is $H = 1.66 g_{*}^{1/2}T^2/m_{\rm pl}$, where
$m_{\rm pl}$ and $g_{*}$ are the Planck mass and the number of matter species,
respectively, 
and $\alpha_{\rm st} = g_{\rm st}^2/4\pi$. 
Hence, above the temperature
$T_{{\rm eq}} \sim g_{*}^{1/2}\alpha_{\rm st}^{2}m_{\rm pl}/1.66 
\sim 3.8 \times 10^{16} \:{\rm GeV}$,
or above the energy density 
$\rho_{{\rm eq}} \sim (1.6 \times 10^{17}\:{\rm GeV})^4$,
the thermalization processes are too slow to maintain equilibrium distributions.
However, in the long-string phase, the rates for thermalization processes
are proportional to 
the densities of the string-bits, $E/V$,
where $E$ and $V$ are the total energy and the volume of the universe, 
respectively.
Since the Hubble expansion rate is proportional to 
the square root of the energy density, 
$\sqrt{E/V}$,
the rates for thermalization processes will dominate
the Hubble rate for  sufficiently high energy density.
Therefore, there will be a critical energy density $\rho_{*}$,
above which equilibrium distributions are realized.\footnote{
Thus the  initial condition $n_{B}=n_{L}=0$ is assured
for sufficiently high energy density.}
Below $\rho_{*}$, interactions freeze out and matter distributions 
are merely affected by the expansion of the universe 
and depart from thermal equilibrium distributions.

Decay processes of higher excited states begin when the energy density decreases
to $\rho_{{\rm decay}} \sim (3.7 \times 10^{17}\:{\rm GeV})^4$,
where the decay rates $\sim \alpha_{\rm st} m_{\rm st}$ dominate the Hubble expansion rate.
These processes are not adiabatic since the matter distributions have departed
from equilibrium ones.

During the decay processes of higher excited states, 
baryon asymmetry as well as entropy is generated.
We can make a rough estimate,
\beq
n_{B}/s \sim \alpha_{\rm st}^2 \sim 10^{-3}, \label{eq:roughdecay}
\eeq
since the first nontrivial contribution to {\it CP} violation comes from
the interference of the lowest-order diagrams and the one-loop corrections.

However, if many processes occur at $\rho =\rho_{{\rm decay}}$,
some cancellations among the decay processes can decrease 
the above result.
These cancellations might be possible 
if many excited states are taken into account,
since ten-dimensional superstring theory has no {\it CP} violation originally.
The cancellations might be explained also since {\it CPT} invariance and 
unitarity assure the relation 
$ \sum_{X} \Gamma (X \rightarrow b) = \sum_{X} \Gamma (X \rightarrow \bar{b})$.

We cannot make a precise estimate since we do not know the dynamics of the decay
processes in detail.
Hence let us consider the following case:
baryon asymmetry is generated after the decay, while 
it is not generated during the decay due to exact cancellations. 
By considering this case, we can give a lower bound on $n_{B}/s$. 
We assume that the following matter distributions are caused by the 
decay processes:
\beq
n_{\phi} = n_{\phi^*} \neq n_{l} = n_{l^*}, \hspace{1cm} (T=T_{\rm H})
\label{eq:initial}
\eeq    
where $n_{\phi}$ and $n_{l}$ are the number densities of the Higgs bosons and 
lepton doublets, respectively.
These nonequilibrium distributions are maintained 
until the temperature of the universe decreases to 
$T_{{\rm eq}}$.
During this epoch a lepton asymmetry is generated.
This lepton asymmetry will be converted into a baryon asymmetry 
of the same order of magnitude 
{\em via} sphaleron processes.
In the next subsection, we estimate the resultant lepton asymmetry
by using the nonequilibrium distribution of Eq.(\ref{eq:initial}) as an 
initial condition.
 
Finally, we make a brief comment on another scenario which causes 
nonequilibrium distributions.
If we assume the inflationary universe, 
inflaton decay processes cause nonequilibrium.
While the inflaton decays, a baryon asymmetry as well as entropy can be generated
\cite{inflation}.
Even if a baryon asymmetry is not generated during the decay processes,
some nonequilibrium distributions like those of Eq.(\ref{eq:initial}) 
are generated.
Then a baryon asymmetry is generated after the inflaton decays.


\subsection{Boltzmann equations} \label{sec:Boltzmann}
\begin{figure}
\begin{center}
\leavevmode
\epsfbox{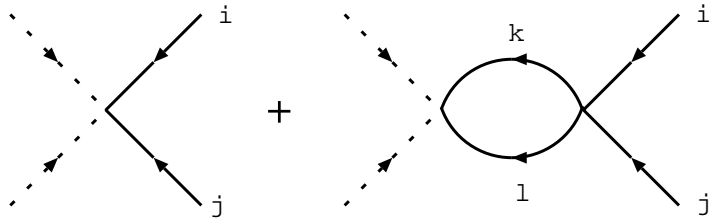}
\caption{Diagrams which contribute to leptogenesis.
The first nontrivial contributions come from the interference between
tree-level amplitudes and one-loop corrections.
The indices $i,j,k$ and $l$ represent the generations.}
\label{fig:lepton}
\end{center}
\end{figure}

\begin{figure}
\begin{center}
\leavevmode
\epsfbox{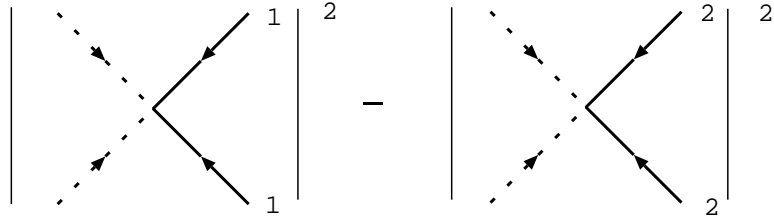}
\caption{Processes which produce 
         a different  number density for generation 1 and generation 2. }
\label{fig:generations}
\end{center}
\end{figure}

In this subsection we calculate the resultant lepton asymmetry  by
using the model of Eq.(\ref{eq:model}) and the nonequilibrium distributions 
of Eq.(\ref{eq:initial}). 
We consider the processes $ll \leftrightarrow \phi^{*}\phi^{*}$ and 
processes related by {\it CP} conjugation.
The first nontrivial contributions to the generation of lepton asymmetry
come from the 
interference terms of the tree-level amplitudes and the one-loop corrections 
shown in Fig. \ref{fig:lepton}.
These are proportional to $\Im (\sum_{k,l} h_{ij}^* C_{ijkl} h_{kl}) \Im (I)$,
where $I$ is a factor coming from the loop integrations.
However, they vanish if we naively sum over the indices for generations 
$i$ and $j$,
since $C_{ijkl}^* = C_{klij}$, as is evident form the Lagrangian of 
Eq.(\ref{eq:model}).
Hence, we consider the processes shown in Fig. \ref{fig:generations}
in order to produce a different number density
for generation 1 and generation 2.

For simplicity, we make the assumption that
the distributions for matter remain  near equilibrium, 
$\rho \sim \exp(\frac{E-\mu}{T})$ and $\mu \ll T$.
Thus the Boltzmann equations for the above processes are as follows:
\beqa
&&\dot{Y_l}+\Gamma_{{\rm th}} (Y_l -1) = 0, \n 
&&\dot{Y_{\phi}}+\Gamma_{{\rm th}} (Y_{\phi} -1) = 0, \label{eq:Boltzmann0}\\ 
&&\dot{Y_{1-2}}+\Gamma_{{\rm th}} Y_{1-2} = 
  \frac{3}{\pi}(\frac{\alpha_{\rm st}}{m_{\rm st}})^2 (h_{11}^2-h_{22}^2)
T^3(Y_{\phi}^2-Y_{l}^2),\label{eq:Boltzmann1}\\ 
&& \dot{Y_{L}} 
= -\frac{72}{\pi} \frac{\alpha_{\rm st}^3}{m_{\rm st}^4}h_{11}h_{22}\Im(C_{1122})
T^5(Y_{l_1}^2-Y_{l_2}^2), \label{eq:Boltzmann2}
\eeqa
where $Y_i = n_i/n^{({\rm eq})}$.
$Y_{1-2}=Y_{l_1}-Y_{l_2}$, $Y_{L}=Y_{l_1}+Y_{l_2}-(Y_{l_1^*}+Y_{l_2^*})$.
Here $\Gamma_{{\rm th}} \approx g_* \alpha_{\rm st}^2 T$ 
is the rate for thermalization.
We use the convention here in which the coefficient of the Majorana-type 
interactions,
$h_{ij}$, is in real diagonal form.

Equations (\ref{eq:Boltzmann0}) represent the thermalization
processes which reduce the nonequilibrium distributions of Eq.(\ref{eq:initial})
imposed as an initial condition to the equilibrium distributions. 
Equation (\ref{eq:Boltzmann1}) represents the processes which produce
the difference in the number densities between the generations.
Finally, the third equation, Eq.(\ref{eq:Boltzmann2}), represents 
the production of a lepton asymmetry.
The right-hand sides of Eqs. (\ref{eq:Boltzmann1}) and (\ref{eq:Boltzmann2})
are given by calculating the amplitudes shown in Fig. \ref{fig:generations}
and Fig. \ref{fig:lepton}, respectively, 
and performing the phase space integrations.

The Boltzmann equations are integrated to give
\beqa
Y_{L}&=& -\frac{36}{\pi^2} \alpha_{\rm st}^{5}
         (\frac{m_{\rm pl}}{1.66g_*^{1/2} m_{\rm st}})^2 (\frac{T_{\rm H}}{m_{\rm st}})^4 J \\
     &\approx & 2.7 \times 10^{-12} J,\label{eq:semifinal}
\eeqa
where
\beqa
J &=& (h_{11}^2-h_{22}^2)h_{11}h_{22}\Im(C_{1122})K, \label{eq:J}\\
K &=& 12 (T_{{\rm eq}}/T_{\rm H})^4 
         \int_{T_{{\rm eq}}/T_{\rm H}}^{\infty}dz z^{-4}(\exp(-z)+A_{l}\exp(-2z))\n
  & &       \times \int_{T_{{\rm eq}}/T_{\rm H}}^z dz' z'^{-2}
          (2(A_{\phi}-A_{l})+(A_{\phi}^2-A_{l}^2)\exp(-z'))\label{eq:integral}\\
  &\approx&  0.538(A_{\phi}-A_{l})+0.173A_{l}(A_{\phi}-A_{l})\n
  & &          +0.108(A_{\phi}^2-A_{l}^2)+0.0355A_{l}(A_{\phi}^2-A_{l}^2) 
             \label{eq:integral2},\\ 
A_i &=& (Y_i (T_{\rm H})-1)  \exp (T_{{\rm eq}}/T_{\rm H}), 
\eeqa
and $T_{{\rm eq}}=g_*^{1/2}\alpha_{\rm st}^2 m_{\rm pl}/1.66 
\sim 3.75 \times 10^{16} \:{\rm GeV}$ is the temperature below which 
thermalization processes begin.
In the estimations of Eqs. (\ref{eq:semifinal}) and (\ref{eq:integral2})
we used the following values:
$\alpha_{\rm st} = 1/45$, $g_{*}=106.75$, $m_{\rm pl}=1.22\times10^{19}\:{\rm GeV}$,
$m_{\rm st}=5.27\times10^{17}\:{\rm GeV}$ and $T_{\rm H}=4.92\times10^{16}\:{\rm GeV}$.

The lepton-to-entropy ratio is 
\beq
n_{L}/s \approx Y_{L}/g_{*} \approx 2.6\times10^{-14} J,
\label{eq:final}
\eeq
where $J$ is given by Eq.(\ref{eq:J}) and its value can be a few thousand
if $h_{ij}$ and $C_{ijkl}$ are about five.
This lepton asymmetry will be converted into baryon asymmetry of the  
same order of magnitude {\em via} sphaleron processes.
Therefore, the observed baryon asymmetry can be generated after the 
decay processes of higher excited states of string theory. 

If we take into account the baryogenesis during the decay processes,
we will obtain a larger value for the baryon-to-entropy ratio.
Too large a value could be diluted afterwards
by considering entropy generation in the confinement-deconfinement
phase transition, for example.
Therefore, the observed baryon-to-entropy ratio can be explained 
in these scenarios.


\subsection{Some comments}\label{sec:comment}

\begin{figure}
\begin{center}
\leavevmode
\epsfbox{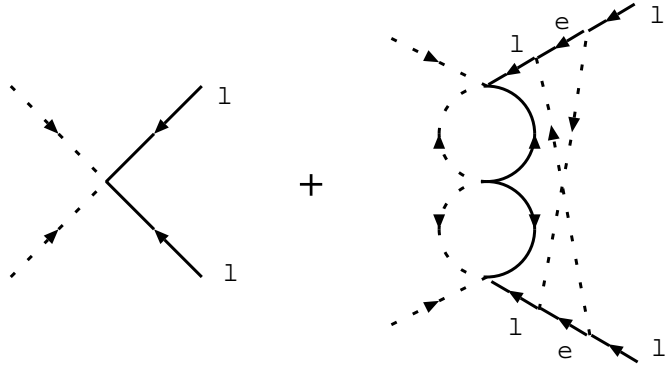}
\caption{Diagrams which contribute to leptogenesis 
by means of Yukawa couplings and Majorana-type couplings.
The left-handed lepton doublets and the right-handed lepton singlets are
$l$ and $e$, respectively.}
\label{fig:yukawa}
\end{center}
\end{figure}

We make some comments on the model of Eq.(\ref{eq:model}).
We have considered Majorana-type interactions plus
four-fermion interactions other than the MSM
in order to introduce {\it CP} violation.
It seems that only Majorana-type interactions would be sufficient since 
there are Yukawa couplings in the MSM already.
The Yukawa plus Majorana-type interactions,
\beq
y_{ij} e^{*\alpha i}l_{\alpha}^{aj} \phi^*_a
+\frac{1}{4} \frac{g_{\rm st}^2}{m_{\rm st}} h_{ij}
 \epsilon^{\alpha \beta}
(\epsilon_{ab} \epsilon_{cd}+\epsilon_{ad} \epsilon_{cb})
 l_{\alpha}^{ia} \phi^b l_{\beta}^{jc} \phi^d
+{\rm h.c.},
\eeq
would also work,
since, after the Yukawa coupling is brought to the form of a  real diagonal 
matrix {\em via} a unitary 
transformation, no degrees of freedom remain to insure a
real Majorana-type coupling.
However, because the Yukawa coupling is small, the resultant lepton 
asymmetry is too small to explain the observed value of $n_{B}/s$.
Indeed, 
a lepton asymmetry is generated, for example, 
through the processes shown in Fig. \ref{fig:yukawa}.
The result is of the following order:
\beqa
n_L/s &\sim& (h\frac{\alpha_{\rm st}}{m_{\rm st}})^{4} y^4 
                      \frac{m_{\rm pl}}{1.66 g_{*}^{1/2}}T_{\rm H}^3 \frac{1}{g_{*}} \n
               &\sim& 6.7 h^4 \times 10^{-21}. 
\eeqa
Here $h$ and $y$ are characteristic values for $h_{ij}$ and $y_{ij}$, 
respectively.
We think $h$ is of the order one, and we used 
the Yukawa coupling of the tau lepton for $y$.

Finally, the Majorana-type terms in Eq.(\ref{eq:model}) give neutrino masses
of the order of\footnote{
Similar results are given in \cite{Fukugita}.}
\beq
m_{\nu} \sim h \frac{g_{\rm st}^2}{m_{\rm st}} v^2 
         \sim 3.2 h \times 10^{-5}\hspace{2mm}{\rm eV}.
\eeq
This value is consistent with the results of the solar neutrino experiments, 
if we consider the vacuum oscillation \cite{HL} or 
take into consideration the magnetic field in the sun \cite{Takasugi},
but it is inconsistent with the results of the atmospheric neutrino 
experiments. 

\section{Conclusions and Discussions}

In this article, we considered the origins of the observed baryon asymmetry
of the universe.
Several plausible origins are possible: 
in the electroweak theory,
the grand unified theory, 
the string theory, {\it etc}.
However, we do not know which origin or which combination of them 
really explain the observed baryon asymmetry.
The origin of the baryon asymmetry has not been established yet.

At the string scale or the Planck scale, the three Sakharov's conditions  
for baryogenesis are satisfied and the observed baryon asymmetry 
 can be generated, as we saw in section \ref{sec:stringbg}.
In order to calculate the precise value, 
we must understand the dynamics of string theory in more detail.

If the baryon asymmetry generated at high temperature vanishes 
due to  washing out by some baryon number
violating interactions in thermal equilibrium, 
or/and dilution by entropy generation in phase transition, 
or/and dilution by inflation of the universe, {\it etc},
it must be generated again at lower temperatures.
For the baryogenesis at the GUT scale and below,
we can use field theory so that more precise estimates are possible.

Within the minimal standard model of the electroweak interactions,
it is difficult to explain the observed baryon asymmetry,
due to the insufficient electroweak phase transition 
and the too small $CP$ violation.
Hence some extensions are necessary. 
Until now, many extended models have been studied and shown to 
generate the observed baryon asymmetry at the electroweak scale.
Hence we cannot specify the model beyond the MSM now.
However, further progress in the understanding of the Higgs sector and 
$CP$ violation
is expected from the near-future experiments:
high energy collider experiments, B-meson experiments, 
measurements of electric dipole moments of
neutron and atoms, {\it etc}.
It will help to establish or rule out the electroweak origin of 
the baryon asymmetry.   

\begin{center} \begin{large}
Acknowledgements
\end{large} \end{center}

I would like to thank K. Hagiwara, who organized the KEK meeting.
I would like to thank K. Funakubo, K. Inoue, H. Kawai, 
Y. Okada and A. Sugamoto for useful discussions.
I am also grateful to Z. Fodor for carefully reading the manuscript.

\end{document}